# Averaged effective pinning potential in YBCO single crystals close to $T_c$

V.Yu. Monarkha, A.G. Sivakov, V.P. Timofeev.

B.Verkin Institute for Low Temperature Physics and Engineering
of the National Academy of Sciences of Ukraine, 47 Lenin Ave., 61103 Kharkiv, Ukraine.

The major part of the publications concerning the magnetic flux dynamics in HTSC are focused on the vortex lattice transformation processes and it's interaction with the crystalline structure of the material. Those processes define the critical current carrying abilities of HTSC in strong magnetic fields [1,2]. On the other hand the low magnetic field range at temperatures close to critical ($T_c$) remains poorly studied. First of all because the magnetic response of the sample is proportional to the field applied, and becomes quite low. Also strong thermal fluctuations in the selected temperature range lead to an increase in the noise level. Both factors make contactless magnetometric study a non trivial task. Despite a big number of theoretical and experimental works on the subject, the full understanding of complex processes of flux pinning and creep is not yet achieved [3].

In this paper we present our results on the low density magnetic flux pinning characteristics, obtained though HTSC samples magnetization relaxation measurements $M(t)$ near the phase transition temperatures (0,8 < $T/T_c$ > 0,99).

As the main object of investigation we have chosen pure YBa$_2$Cu$_3$O$_{7-x}$ (YBCO) single crystal samples with optimal oxygen saturation. The dimensions of the samples were close to 1x1 mm$^2$ (in *ab* plane) and 0.02 mm in thickness (along the *c* axis). All samples contained unidirectional twin boundaries parallel to the *c* axis through all the volume. Twin boundaries include CuO$_x$ layers containing oxygen vacancies and dislocations along the twinning plane. These defects locally suppress the superconductive order parameter creating effective pinning cites.

The contactless squid-magnetometric method provided good susceptibility in the selected field and temperature range. The temperature stability during an isothermal relaxation measurement was ~5 mK in 50 – 95 K temperature range.

The residual magnetic moment is defined by the trapped fields in the volume of the sample during FC (Field Cooling). In this case the influence of the surface barriers on the magnetic flux dynamics is minimal [1,2]. The data obtained from the magnetic flux dynamics analysis can be used to define some important pinning parameters. For example, the effective pinning depth potential ($U_p$). In the simplest case it can be estimated from normalized isothermal relaxation rate $S$ using Anderson-Kim linear flux creep model:

$$S = 1/M_0 \, (dM/d\ln t) = - k \, T/U_p. \qquad (1)$$

Most of the published works deal with strong magnetic fields (~kOe) where the vortex lattice is well formed and the pinning parameters are defined by significant vortex-vortex interactions.

Magnetization relaxation curves, normalized on the initial value $M_0$ are presented on Fig.1. The homogeneous magnetic field of the solenoid was parallel to the *c* axis of the sample and was equal to 160 A/m (~ 2 Oe). This field orientation provides optimal vortex pinning on twin boundaries. When the temperature gets close to $T_c$ strong thermal fluctuations cause giant vortex and vortex bundle creep, and in fact vortex liquid state can be observed. In the samples under test, full flux exit (residual magnetization drop to null) has been observed at 91.4 K. The insert on Fig.1 shows a typical superconductive phase transition curve of the sample.

Applying the Anderson-Kim model to the magnetization relaxation curves obtained, we built a temperature dependence plot of the volume averaged effective pinning potential (Fig.2). In comparison to previously known works (for example [2,4]) this temperature and field range has never been studied so closely before, and new regularities have been revealed. First of all, a significant increase of the effective pinning potential value is observed. It can reach up to a few decades of eV on the initial part of the temperature range. Also significant nonmonotonicity of the $U_p(T)$ dependence is observed. In these conditions it can't be explained by the vortex lattice structure reorganization [5].

The unexpected effective pinning potential $U_p$ increase observed can be connected with the dependence of the pinning potential on the applied magnetic field and induced currents, which is poorly studied [6]. Low constant fields and thus low trapped flux density allowed us to study the initial part of the $U_p(H, J)$, where nonlinear models assume existence of a maximum of pinning potential [1,2].

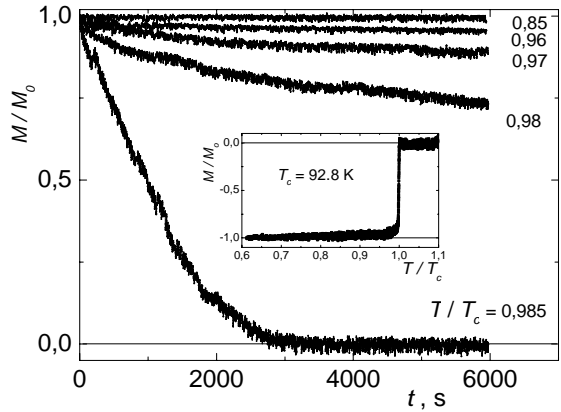

Fig. 1. Isothermal magnetization relaxation curves $M(t)/M_0$ of one of the YBCO single crystal samples under test with unidirectional twin boundaries at different temperatures. The inset shows a typical superconductive phase transition of the sample.

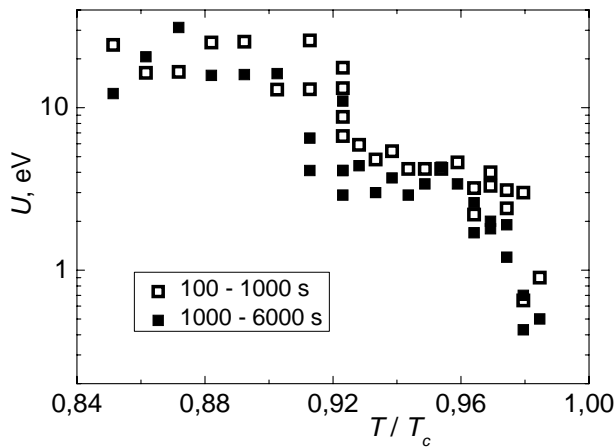

Fig. 2. The temperature dependence of the effective pinning potential. Different markers correspond to different time frames used to calculate $U_p$.

The nonmonotonic decrease of the effective pinning observed close to $T_c$ can be connected with the interplay between thermoactivation energy growth, and the potential depth increase of the pinning cites.

The critical current density, one of the most important parameters of HTSC, is mainly defined by the averaged pinning potential $U_p$. The estimation of the critical current value can be performed both resistively and using contactless magnetometric methods. In order to compare the data obtained from our magnetization relaxation measurements, direct resistive and magnetization loop measurements were performed. The magnetization loop width is proportional the effective pinning force [2]. In the temperature and field range under study the data obtained using thee different methods agree qualitatively.

In conclusion we have demonstrated our results on magnetic flux dynamics trapped in low magnetic fields at temperatures close to $T_c$ in YBCO single crystal samples with unidirectional twin boundaries. We note the presence of strong pinning in the system of unidirectional planar defects and a significant nonmonotonicity of the $U_p(T)$ dependence behavior for low density trapped magnetic flux.

The authors are grateful to the Low temperature physics department of V. N. Karazin Kharkiv National University for the samples provided, and to Yu. Savina for her indispensable assistance.